\documentclass[a4paper]{jpconf}
\usepackage{graphicx}
\usepackage{tabularx}
\usepackage{amsmath}

\usepackage{iopams}
\usepackage[sort&compress,numbers,square,comma]{natbib}

\newcolumntype{M}[1]{>{\centering\arraybackslash}m{#1}}
\newcommand{\exy}{\ensuremath{E_{x,y}}}
\newcommand{\ex}{\ensuremath{E_{x}}}
\newcommand{\ey}{\ensuremath{E_{y}}}

\newcommand{\aest}{\ensuremath{\alpha}}
\newcommand{\best}{\ensuremath{\beta}}
\newcommand{\aens}{\ensuremath{\boldsymbol{\alpha}}}
\newcommand{\bens}{\ensuremath{\boldsymbol{\beta}}}

\begin{document}
\title{Ensemble Estimation of Network Parameters: A Tool to Improve the Real-time Estimation of GICs in the South African Power Network}
\author{M.J. Heyns$^{1,2}$, S.I. Lotz$^{1}$, P.J. Cilliers$^{1}$, C.T. Gaunt$^{2}$}
\address{$^{1}$South African National Space Agency, Space Science Directorate, Hermanus, South Africa\\
$^{2}$Department of Electrical Engineering, University of Cape Town, Cape Town, South Africa}
\ead{mheyns@sansa.org.za}
\begin{abstract}
It has long been known that large grounded conducting networks on the surface of Earth, such as power lines or pipelines, are affected by solar activity and subsequent geomagnetic storms. With the increased use of electrical technologies, society has become more and more dependent on electrical power and power networks. These power networks form extensive grounded conductors which are susceptible to geomagnetically induced currents (GICs). GICs at any specific node in a power network are assumed to be linearly related to the horizontal vector components of an induced plane-wave geoelectric field by a pair of network parameters. These network parameters are not easily measured in the network, but may be estimated empirically. In this work, we present a new approach of using an ensemble of network parameters estimates. The ensembles include a huge number of parameter pair estimates calculated from simultaneously solving pairs of time instances of the governing GIC equation. Each individual estimate is not the true state of the system, but a possible state. Taking the ensemble as a whole though gives the most probable parameter estimate. The most probable parameter estimate for both network parameters, as defined by their respective ensembles, is used directly in the modelling of GICs. The ensembles themselves however allow for further analysis into the nature of GICs. An improvement is seen when comparing the out-of-sample performance of the ensemble estimates with previous GIC modelling in the South African power network during the Halloween Storm of 2003. For the first time, it is shown that errors in the GIC modelling chain are absorbed into the network parameter estimates. Using a range of estimates from the ensemble, a GIC prediction band is produced. This band corresponds to an error estimate for predicted GIC. Furthermore, it has been explicitly shown for the first time that estimated network parameters vary with GIC magnitude during an event. This behaviour is then used to refine the parameter estimation further and allow for real time dynamic network parameter estimation. 
\end{abstract}
\vspace*{-1cm}
\section{Introduction}
Modelling geomagnetically induced currents (GICs) entails not just modelling a single system, but rather a chain of coupled systems. The general driver of a geomagnetic storm disrupts current systems in the near-Earth environment and hence the geomagnetic field. Fluctuations in the geomagnetic field through a closed path including the power network induces a geoelectric field and an electromotive force (EMF), which ultimately drives the GIC. The effects of the resulting GICs in these networks can be particularly damaging and costly \cite{Oughton2017}, especially in high-latitude regions. Lesser, but still significant effects have also been observed at mid-latitudes \cite{Koen2002a,Kappenman2003,Gaunt2007}.

There are a few governing assumptions made in GIC modelling. It is assumed that the induced geoelectric field is driven by a spatially constant plane-wave geomagnetic field over the system \cite{Pirjola1982}. Such an induced geoelectric field would be uniform if the conductivity of Earth is dependent only on depth (laterally constant). Typical causes of such a geomagnetic field would be a very long (relative to the system) uniform sheet or line current. These assumptions are idealised cases which can be thought of as first order approximations \cite{Boteler2017}. In reality, the ionospheric current systems are exceptionally complex, non-uniform and highly variable. At mid-latitudes the useful plane-wave geomagnetic field assumption is more accurate than in auroral regions where the auroral electrojet is the dominant current system. This current system is more variable and closer compared to the mid-latitude drivers, making the interactions more complicated \cite{Boteler1998a,Kappenman2003,Wik2008}.
\vspace*{-15pt}\subsection{Network Parameters}
Due to the conductivity and extent of the Earth, the main component of the driving EMF comes from the induced geoelectric field at the surface of the Earth. This results in a linear relation between the geoelectric field and the resulting GIC through Ohm's law, defined by the projection of the effective geoelectric field onto the network and the sum of all resistances in the induction loop (which is dominated by network resistances). This results in the governing GIC equation in Cartesian coordinates,
\begin{equation}
\label{eqn:gic}
GIC(t)=a\ex(t) + b \ey(t),
\end{equation}
where \ex\ and \ey\ are the horizontal geoelectric field components in the Northerly and Easterly directions respectively. Since the network parameters are scaling factors that also penalise non-alignment of the geoelectric field vector with the network, the preferred direction for the geoelectric field to produce large GICs can be found. The ratio of network parameters $b/a$ as used in previous work \cite{Ngwira2008} is encoded in this direction (in this case effectively the angle from North). Any deviation or non-alignment of the geoelectric field from the preferred direction would result in a fraction of the total geoelectric field magnitude contributing to the measured GIC. These traditional network parameters, with units of Akm/V, are usually determined analytically using available information about the network, which is often inaccurate. 
\vspace*{-10pt}\section{Model Development}
From simultaneous GIC and \exy\ data, a single estimate of the network parameters can be derived empirically \cite{Wik2008,Matandirotya2015}. It is assumed that these parameters are constant over the time scale of a geomagnetic event, only changing with major alterations of power network related hardware or operations \cite{Viljanen2013}. Due to associated errors at each point in the GIC modelling chain, it has been observed that different empirical values of $a$ and $b$ may be derived for different data subsets defined by magnitude or conductivity \cite{Wik2008,Matandirotya2015}. This suggests that a number of errors from different drivers are absorbed into these parameters. To acknowledge these errors and borrowing notation from Wik et al. \cite{Wik2008}, where \aest\ and \best\ represent the empirically derived network parameters, we can redefine the governing equation to,
\begin{subequations}\label{eqn:newgic}
\begin{equation}
\Gamma(t)\approx\alpha\ex(t)+\beta\ey(t) \mbox{, where} \tag{\ref{eqn:newgic}}
\end{equation}
\vspace*{-20pt}
\begin{equation}
\Gamma(t) \equiv GIC(t)+GIC(t)_{err} \mbox{ (or the GIC as measured),} \label{eqn:newgica}
\end{equation}
\vspace*{-18pt}
\begin{equation}
\aest\equiv a\big(1+{\ex}(t)_{err}/{\ex(t)}\big) \mbox{ and} \label{eqn:newgicb}
\end{equation}
\vspace*{-18pt}
\begin{equation}
\best\equiv b\big(1+{\ey}(t)_{err}/{\ey(t)}\big). \label{eqn:newgicc}
\end{equation}
\end{subequations}
Given a time-series of $n$ time instances, each a measured state of the system, taking any two relevant and comparable time instances would allow for \aest\ and \best\ to be solved for simultaneously. Using all possible combinations of time instance pairs would result in $n(n-1)/2 \approx n^2/2$ (for large $n$) sets of empirical network parameter estimates. These estimates are collected into the ensembles \aens\ and \bens. Using a similar approach of statistical mechanical ensembles, each estimate in the ensemble represents a possible state of the real system. When all states are considered together and the result normalised appropriately, an ensemble becomes a probability distribution of system states.

Previous empirical estimation has generally included a fit of sorts that results in a single estimate. One approach is to use a least squares routine and fit equation (\ref{eqn:gic}) to the data \cite{Wik2008}. Another approach is to select near-zero crossings of the geoelectric field for a single geoelectric field component and solve equation (\ref{eqn:gic}). This generates a number of single parameter estimates and a linear fit is then applied (excluding outliers) to find the final estimate \cite{Matandirotya2015}. It has also been shown that the ratio of network parameters can be found empirically from data \cite{Pulkkinen2007}. If a single parameter is known, this ratio can be used to find the other parameter in the parameter pair \cite{Ngwira2008}. All these methods estimate the network parameters and absorb effects analytical methods don't, but do not give an indication of any variation in the parameters (in which errors are propagated), whereas the spread of the resulting ensembles do. 
\vspace*{-10pt}\subsection{Data Sources and Selection}
This work analyses the Grassridge (GRS) substation in the South African power network, where previous GIC modelling has been done. Measured 2-sec transformer neutral GIC data for 31 March 2001 and 29-31 October 2003 is used. This data range spans geomagnetic storms and no network changes were made during this time (comparable time instances). The geoelectric field data used in this work is derived from 1-min magnetometer data measured at Hermanus (which has been shown to be `local' enough for GIC modelling \cite{Ngwira2009}). This process makes use of the magnetotelluric method with a layered Earth conductivity model \cite{Pulkkinen2007}. The conductivity profiles used in this work include the local empirically derived 10-layer profile for Grassridge \cite{Ngwira2008} and the non-local 5-layer for Qu\'{e}bec (QUE) \cite{Boteler1998b}. To make use of relevant data, a process of data selection was implemented. In previous work this entailed using only GIC data which satisfies $|GIC|>0.1\times RMS(GIC)$, where $RMS=\sqrt{{\Sigma^n_{i=1}GIC(t_i)^2}/{n}}$ (root mean square), and further selection of significant geoelectric field time instances (which can lead to biasing). Since ensemble estimation is robust and makes use of all possible combinations of time instances, this selection can be relaxed and varied, with the only criterion being that there need to be enough time instances to create a large representative ensemble.
\vspace*{-10pt}\section{Ensemble Estimation Results}
\begin{figure}[h]
\includegraphics[width=0.6\textwidth]{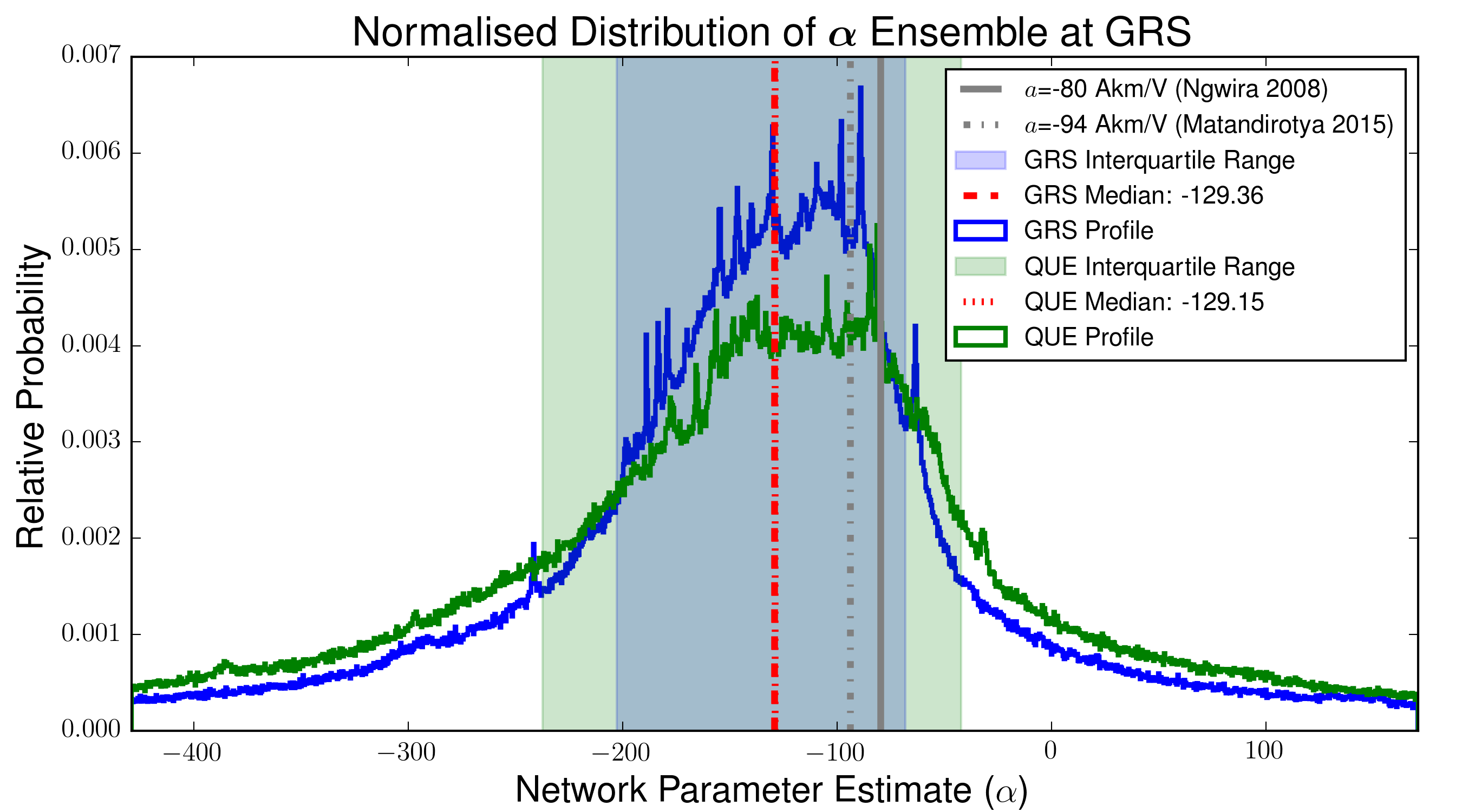}\hspace{0.01\textwidth}%
\begin{minipage}[b]{0.39\textwidth}
\caption{\label{fig:adis} Comparison of the \aens\ distributions derived from the local (GRS) and Qu\'{e}bec (QUE) conductivity profiles. This behaviour also holds for the \bens\ distributions. The interquartile range shown is used to model a prediction band (Figure \ref{fig:band}). Also shown are the network parameter estimates from previous work.} 
\vspace*{0.75cm}
\end{minipage}
\vspace*{-20pt}
\end{figure}

In order to compare ensemble estimation to previous work the same GIC data selection criterion is used initially. The data from the Halloween Storm of 29 October 2003 is kept out of the ensemble training as a validation set (not done in previous work). Figure \ref{fig:adis} shows the resulting normalised \aens\ ensemble histogram (with a similar result applicable for \bens). It should be noted that although the probability distribution has the expected bell-shape as predicted from equation (\ref{eqn:newgicb}), it also has significantly heavy tails (variance and higher order moments are undefined). In this case, the median and not the mean is the best measure of central tendency and hence the most probable network parameter estimate. Also shown is the effect of using different conductivity profiles, with the local empirically derived profile having less spread but no shifting of the peak. This shows that errors made in the derivation of the geoelectric field are absorbed into the empirical network parameters.
\vspace*{-10pt}\subsection{Comparison with Previous Work}
A comparison is made with the empirical network parameter estimates of two previous studies, namely Ngwira et al. \cite{Ngwira2008} and Matandirotya et al. \cite{Matandirotya2015}. Both these studies used the first day of the Halloween Storm, 29 October 2003, for GIC modelling. During this day, the period between 06:00 and 12:00 is highly disturbed according to SYM-H and the period between 19:00 and 24:00 is relatively disturbed (see Figure \ref{fig:band}). The periods 06:00-12:00 and 19:00-24:00, will be referred to as the Primary and Secondary Storm Phases respectively. In addition to this, Matandirotya et al. \cite{Matandirotya2015} used the finite element method (FEM) to further improve the derived geoelectric field results (not done in this work). 

To measure the performance of the modelling a number of error metrics are used, namely the $RMSE$ (root mean square error), $RE$ (relative error) and Pearson's correlation coefficient $\rho$. The first two metrics quantify the error in absolute amplitude and are defined by, 
\\\begin{minipage}[b]{0.5\textwidth}
$$RMSE=\sqrt{\frac{\sum^n_{i=1}(GIC_{obs}(t_i)-GIC_{mod}(t_i))^2}{n}}$$
\end{minipage}%
\begin{minipage}[b]{0.15\textwidth}
\centering and
\end{minipage}%
\begin{minipage}[b]{0.35\textwidth}
\begin{equation}
RE=\frac{GIC_{obs}-GIC_{mod}}{GIC_{obs}}.
\end{equation}
\end{minipage}
\vspace*{0.1cm}\\Using the previous work's definition, only the median $RE$ for $|GIC|>1$ A is considered and the result is shown as a percentage \cite{Matandirotya2015}. Pearson's correlation coefficient on the other hand is defined in the normal way and quantifies the correlation. The results summarised in Table \ref{tab:results} below show that ensemble estimation does significantly better in general (an overall improvement of more than 10\% using the locally based conductivity profile and more than 5\% using the non-local profile).
\begin{table}[h]
\vspace*{-10pt}
\small
\centering
\caption{\label{tab:results} Ensemble estimation results compared to previous work.}
\makebox[\textwidth][c]{\begin{tabular}{M{2.5cm}M{2.5cm}M{2.5cm}M{2.5cm}M{1.5cm}}
\br
Data& \phantom{RMSE [A] ($\rho$)} 06:00-12:00 &RMSE [A] ($\rho$) 19:00-24:00 & \phantom{RMSE [A] ($\rho$)} 00:00-24:00 &RE\%\\ 
\mr
\multicolumn{5}{l}{\it Ngwira Set (a=-80, b=1 A km/V)}\\ 
 
FEM & 0.96 & 1.07 & 1.35 & 51\\ 
 
\multicolumn{5}{l}{\it Matandirotya Set (a=-94, b=24 A km/V)}\\ 
 
FEM & 1.38 & 1.11 & 0.98 & 41\\ 
 
\multicolumn{5}{l}{\it Grassridge Profile ($\alpha$=-129.36, $\beta$=7.90 A km/V)}\\ 

GRS & \bf 1.42 (0.88) & \bf 0.54 (0.97) & \bf 0.86 (0.88) & \bf 30\\ 

\multicolumn{5}{l}{\it Qu\'{e}bec Profile ($\alpha$=-129.15, $\beta$=5.61 A km/V)}\\ 
 
QUE & 1.78 (0.78) & 0.81 (0.93) & 1.12 (0.79) & 35\\ 
\br
\end{tabular}}
\vspace*{-10pt}
\end{table}
\vspace*{-10pt}\subsection{Network Parameter Dependence on GIC Magnitude}
\begin{figure}[h]
\includegraphics[width=0.7\textwidth]{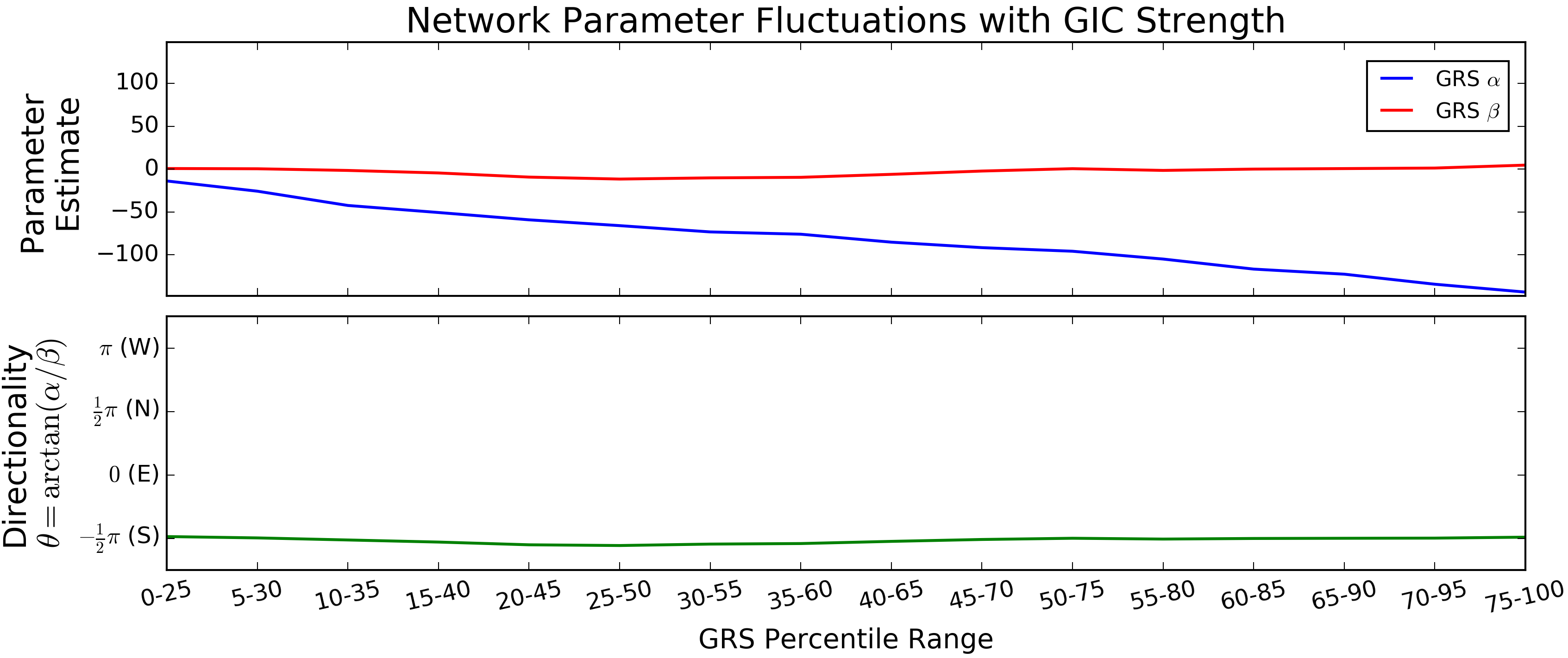}\hspace{0.01\textwidth}%
\begin{minipage}[b]{0.29\textwidth}
\caption{\label{fig:parvgic} Plot of empirical network parameter estimates at GRS for different magnitude defined GIC percentile ranges. Also shown is the constant ratio of network parameters (network defined directionality).}
\vspace*{0.7cm}
\end{minipage}
\vspace*{-20pt}
\end{figure}

Relaxing the criterion for using significant GIC data (considering that already only data corresponding to geomagnetic storms is used), we can use different percentile windows relating to GIC magnitude to quantify the previously observed behaviour that different empirical network parameters are obtained for different data subsets. Taking a 25\% percentile window for GRS results in more than 1 million parameters pairs per window, satisfying the condition that a representative ensemble is produced. Arranging the percentile range estimates along increasing values of measured GIC in Figure \ref{fig:parvgic} shows that \aest\ and \best\ are apparently not constant and vary with GIC intensity as the storm evolves (since GIC magnitude varies with time). At small magnitude GICs, the network parameters are negligible. This suggests that the preferred direction is indeterminate and that these cases are often a result of miss-alignment and/or a small geoelectric field. As the GIC strength grows, the network parameters become more and more relevant. This is not the case with the \best\ parameter at GRS which stays close to zero. This is a function of the directionality weighting from the network. GRS is located at a stable endpoint in the network with only a single, approximately North-South directed line. Therefore the \best\ parameter, which scales the Eastward geoelectric field component \ey, is small. This suggests that GICs at GRS are practically independent of the Eastward component of the geoelectric field, which has been seen in previous analysis \cite{Koen2002a,Ngwira2008,Matandirotya2015}. Even with this variation, the preferred direction as defined by the ratio of network parameters remains constant.
\vspace*{-10pt}\subsection{GIC Prediction Band}
\begin{figure}[h]
\centering
\includegraphics[width=0.9\textwidth]{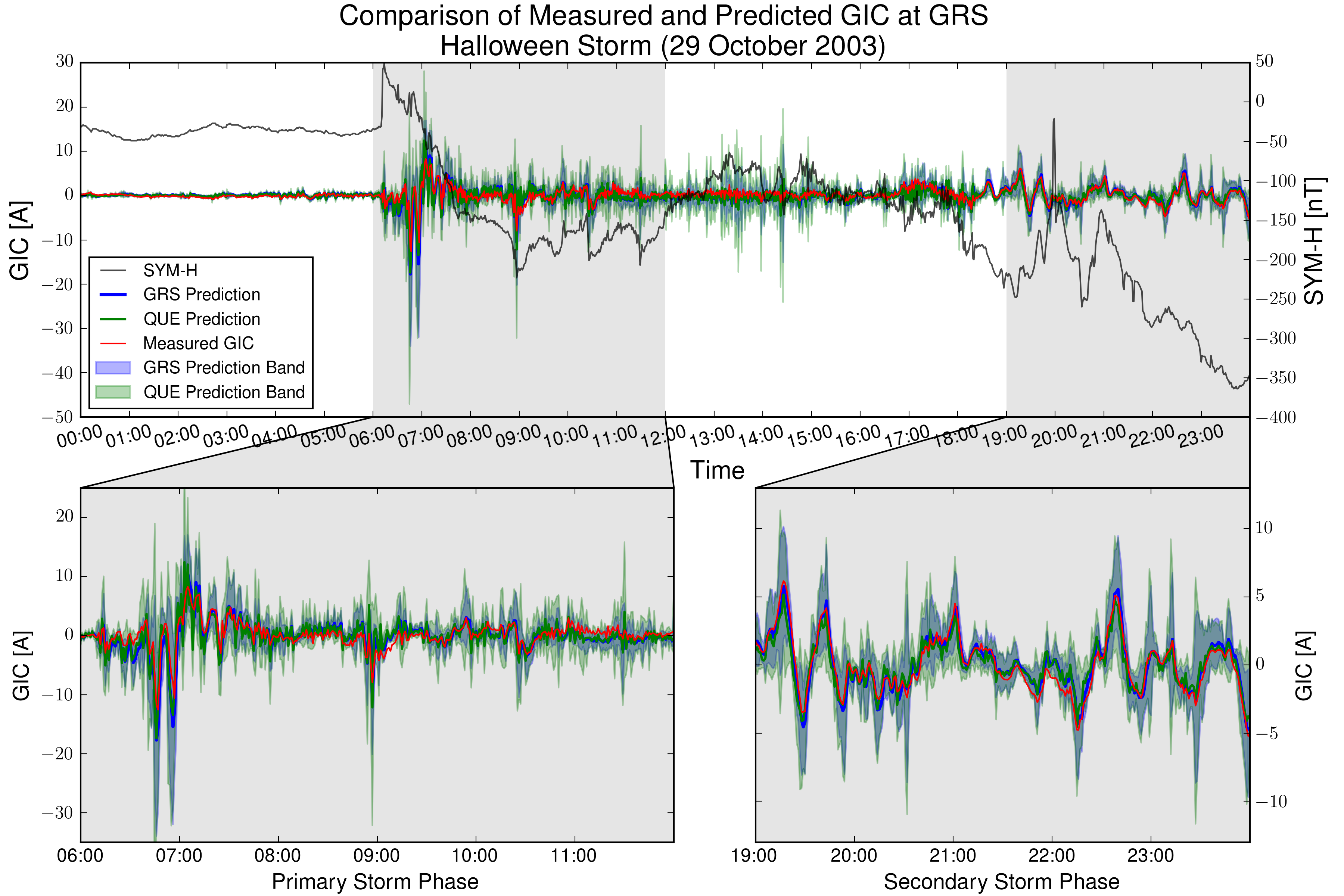}
\vspace*{-10pt}
\caption{\label{fig:band} A GIC prediction band produced using a range of estimates.}
\vspace*{-10pt}
\end{figure}

The spread in the distributions of estimated network parameters suggests that using single values of \aest\ and \best\ to relate GIC to induced geoelectric field is not correct. Instead of using this approach, we can use the interquartile range of \aens\ and \bens\ (see Figure \ref{fig:adis}) to predict an associated range of GICs. This network parameter range would span the typical error propagated in the GIC modelling chain, without straying into the heavy tails. Since we are most concerned about the largest GICs, we then use the values of \aest\ and \best\ in this interquartile range that will maximise the range of resulting GIC. This approach results in a GIC band, as shown in Figure \ref{fig:band}, instead of a single estimate. This gives a handle on the uncertainty in the prediction at any time.
\vspace*{-10pt}\subsection{Dynamic Network Parameters}
Using the observed correlation of the network parameters with GIC magnitude, dynamic network parameters can be created to further improve modelling. The percentile window shifted through all the percentiles results in an overlap of estimates for a given GIC magnitude. Using a weighted mean of these overlapping estimates would result in a single representative estimate for a corresponding GIC magnitude. Calculating an arbitrary GIC magnitude's relevant percentile in the ensemble would then allow the representative network parameter estimate to be mapped to it. Using the GRS data with its low resolution and the non-local QUE profile (that should do a lot worse than a locally derived profile) we see an improvement nonetheless (see Table \ref{tab:dyn}). Also shown are the highest percentile range parameters, appropriate for extreme value modelling.
\begin{table}[h]
\vspace*{-10pt}
\small
\centering
\caption{\label{tab:dyn} Dynamic network parameter estimation using GIC magnitude.} 
\begin{tabular}{M{3.6cm}M{2cm}M{2.4cm}M{2cm}M{0.8cm}}
\br
Qu\'{e}bec Profile& \phantom{RMSE [A] ($\rho$)} 06:00-12:00 &RMSE [A] ($\rho$) 19:00-24:00 & \phantom{RMSE [A] ($\rho$)} 00:00-24:00 &RE\%\\
\mr
Static Parameters &  1.78 (0.78) & 0.81 (0.93) & 1.12 (0.79) & 35\\ 

Dynamic Parameters & 1.75 (0.82) & 0.72 (0.94) & 1.05 (0.82) & 36\\ 
 
Extreme Parameters & 1.97 (0.79) & 0.78 (0.93) & 1.21 (0.79) & 36\\  
\br
\end{tabular}
\vspace*{-10pt}
\end{table}
\vspace*{-10pt}\section{Discussion and Conclusion}
Not only does ensemble estimation give a much better prediction compared to previous analytical and empirical methods, but the uncertainty in GIC modelling is also quantified. It has been shown how errors made in the derivation of the geoelectric field components are reflected in the network parameters. Taking this into account and a range of values for each of the network parameters, a band of GIC may be calculated instead of a single estimate. Further analysis has also explicitly shown for the first time that the network parameters are not constant and are correlated with GIC strength and hence storm phase. Using this to define dynamic or extreme parameter estimates further improves GIC modelling accuracy and usefulness. These approaches can be incorporated into a real-time probabilistic prediction scheme, with direct application to mitigation schemes for power utilities and other relevant parties.
\bibliographystyle{iopart-num}
\vspace*{-10pt}
\bibliography{proceedings}
\end{document}